\documentclass[11pt,a4paper,iop]{emulateapj}

\newcommand\T{\rule{0pt}{2.6ex}}
\newcommand\B{\rule[-1.2ex]{0pt}{0pt}}

\usepackage{epsfig}
\usepackage{amsmath}
\usepackage{natbib}

\shorttitle{}
\shortauthors{Isella A., P\'erez L., and Carpenter J.}

\begin{document}


\title{On the nature of the transition disk around LkCa~15} 
 

\author{Andrea Isella, Laura M. P\'erez, John M. Carpenter}
\affil{Department of Astronomy, California Institute of Technology, MC 249-17, Pasadena, CA 91125, USA}
\email{isella@astro.caltech.edu}

\begin{abstract}
We present CARMA 1.3 mm continuum observations of the T Tauri star LkCa~15,
which resolve the circumstellar dust continuum emission on angular scales between 
0.2\arcsec-3\arcsec, corresponding to 28-420 AU at the distance of the star. The 
observations resolve the inner gap in the dust emission and reveal an asymmetric dust distribution in the outer
disk. By comparing the observations with theoretical disk models, we calculate that 
90\% of the dust emission arises from an azimuthally symmetric ring that contains about 
$5\times10^{-4}$ M$_{\sun}$ of dust. 
A low surface-brightness tail that extends to the northwest out to a radius of 
about 300 AU contains the remaining 10\% of the observed continuum emission. 
The ring is modeled with a rather flat surface density profile between 40 and 120 AU, 
while the inner cavity is consistent with either a sharp drop of the 1.3 mm dust optical depth at about
42~AU or a smooth inward decrease between 3 and 85 AU. We show that early science ALMA 
observations will be able to disentangle these two scenarios.  Within 40 AU, the observations 
constrain the amount of dust between $10^{-6}$ and 7 Earth masses, where the minimum and 
maximum limits are set by the near-infrared SED modeling and by the millimeter-wave observations 
of the dust emission respectively.  
In addition, we confirm  the discrepancy in the outer disk radius inferred from the dust and gas, 
which corresponds to 150 AU and 900 AU respectively.  We cannot reconcile this difference 
by adopting an exponentially tapered surface density profile 
as suggested for other systems, but we instead suggest that the gas surface density in the outer disk 
decreases less steeply than that predicted by model fits to the dust continuum emission. 
The lack of continuum emission at radii lager than 120 AU suggests a drop of at least a factor of 5 
in the dust-to-gas ratio, or in the dust opacity. We show that a sharp dust opacity drop of this magnitude 
is consistent  with a radial variation of the grain size distribution as predicted by existing grain 
growth models.
\end{abstract}



\section{Introduction}
\label{sec:intro}


Over the past decade, the {\it Spitzer Space Telescope} has identified an 
increasing population of circumstellar disks surrounding pre-main sequence 
stars that display a deficit of near- and mid-infrared flux
compared to a ``classical'' T Tauri star, while exhibiting similar levels of
far-infrared emission. The spectral 
energy distribution (SED) of these objects suggests that these stars lack warm 
dust in the inner disks regions, while substantial emission at longer wavelengths 
indicates that the outer disk remains quite massive \citep[see, e.g.,][]{Calvet05}. These 
so-called transition disks may be intermediate systems between 
young ($\sim 1-10$~Myr), optically thick, gas rich protoplanetary disks and 
old ($>10$~Myr), optically thin, gas poor, debris disks, where significant disk 
evolution has occurred and a planetary system might have formed.

Subsequent high-resolution images of the optically thin millimeter-wave dust 
emission confirm the SED modeling predictions that the inner 
disk in transition disks is depleted of dust \citep[][]{Andrews11a,Brown09,Hughes09, Pietu06}.  
However, spatially resolved observations of the gas emission show 
that some of these dust depleted cavities contain a significant amount of gas \citep{Pietu07} and 
that  the size of the hole is not always consistent with SED modeling \citep{Andrews11a}.
Furthermore, recent high angular resolution observations of the mm-wave dust emission have 
also revealed that  some ``classical'' disks have dust depleted inner regions 
similar to those of ``transition'' disks, but do not show any deficit in the infrared 
excess \citep{Isella10a,Isella10b,Andrews11a}. These results demonstrate that SEDs 
alone cannot provide a complete picture of the disk structure. High angular resolution 
millimeter-wave observations are essential to remove some of the degeneracies 
that limit SED modeling, and therefore provide a more accurate description of the radial distribution 
of circumstellar material.

To determine the physical pheno\-mena responsible for disk cavities, 
measurements of the mass surface density profile $\Sigma(r)$ are essential. 
To first order, circumstellar disks evolve via viscous spreading, where friction forces 
redistribute the material and smoothly shape the disk surface density \citep{Lynden74}. 
Disk photoevaporation by energetic photons might reduce the amount of 
disk material. Models that combine these two process have been the focus of many 
theoretical studies \citep{Alexander06a,Alexander06b,Gorti09,Owen10}. 
These models suggest that viscous evolution proceeds 
unperturbed by photoevaporation until the mass accretion rate throughout the disk is 
comparable with the photoevaporation rate. After this phase, which can last for a few Myr 
for typical values of the stellar radiation and disk surface density, a gap opens at few AU from 
the central star, where the photoevaporation rate is the largest. The material within the gap then 
drains quickly onto the central star leaving a gas-free cavity in the disk. Although 
the exact time scale for the formation of the cavity varies between different models,  
they all predict an order of 
magnitude drop in the gas surface density over a 1 AU radial extent at the edge of 
the cavity.

Alternatively, cavities in the disk can be produced by the gravitational interaction 
between giant planets and the disk material \citep[see, e.g.,][]{Bryden99}. Although this is
a common explanation for transition disks, \cite{Zhu11} have shown 
that even multiple giant planets cannot account for optically thin 
cavities larger than 20 AU, unless the small grains within the region perturbed by the 
planets are efficiently removed. Similarly to the photoevaporation model, the planet-disk 
interaction predict a sharp drop of the dusty and gas density at the edge of the region 
cleared by the planet tidal interaction.

A third possible origin of the inner disk cavities is that the dust grains have grown 
from micron sized  particles to centimeter sized pebbles. Therefore, the cavities reflect a 
reduced dust opacity and not a reduced matter density. Theory predicts that dust coagulation is 
more efficient in the dense inner disk \citep{Tanaka05,Dullemond05} and as the dust 
particles grow, the opacity of the dust diminishes. Current models of grain growth anticipate 
a continuous radial variation of the dust opacity within the disk \citep[e.g.,][]{Birnstiel10}, 
whose observational signature corres\-ponds to a gradual transition between the inner and 
outer disk structure. 

Among the known stars with a transition disk, the 2-5 Myr old star LkCa~15 represents an outstanding 
case. LkCa~15 is a K5 star \citep[$L_\star = 0.74 L_\sun$, $\textrm{M}_\star = 1.0 \textrm{M}_\sun$;][]{Simon00,Kenyon95}
located in the Taurus star-forming region at a distance of about 140 pc \citep{vandenAncker98}. 
Its disk is characterized by a partially dust-depleted inner region of 
about 50 AU in radius \citep{Andrews11a,Espaillat07,Pietu06} and a mass accretion 
rate of about  $3 \times 10^{-9}$  M$_\sun$  yr$^{-1}$ \citep{Hartmann98}. 
The disk SED is characterized by the presence of an infrared excess over the stellar photosphere, 
which can be explained by the presence of hot dust 
within few AUs from the central star \citep{Espaillat07,Mulders10}. The disk has a very 
active chemistry observed in several molecular transitions \citep{Chapillon08,Pietu07,Oberg10}, 
and also exibits  keplerian rotation out to a radius of 900 AU \citep{Pietu07}. These latter 
observations suggest that gas is present at an orbital radius as small as 10 AU,
and therefore well within the cavity observed in the dust continuum emission. Ground 
based H-band (1.6~$\mu$m) observations 
reveal the presence of an elliptical structure in correspondence to the 
edge of the dust inner cavity \citep{Thalmann10}. Finally, \cite{Kraus11} has recently reported the 
discovery through aperture masking of a possible 6~M$_J$ mass proto-planet located inside the
continuum cavity at an orbital distance of 16 AU from the central star. These observations
suggest that the cavity might be indeed shaped by the dynamical interaction with a giant 
planet. However, \cite{Zhu11} and \cite{Dodson11} have argued that a
cavity 50 AU in radius cannot be explained by a single planet, and that additional giant planets 
or clearing mechanisms are required to explain these observations.
 
In this paper we present new observations of the 1.3 mm dust thermal emission obtained 
 with the Combined Array for Research in Millimeter-wave Astronomy (CARMA).
The observations achieve an angular resolution of 0.2\arcsec\, corresponding 
to a spatial scale of 28 AU at the distance of star, which is a factor of two 
improvement with respect of existing observation of this system. In Section~\ref{sec:obs} 
and \ref{sec:morph} we present our observations of LkCa~15 and discuss the morphology 
of the dust emission. The method used to analyze the data is discussed in Section~\ref{sec:model}, 
and the results are presented in \ref{sec:res_mod}. In Section~\ref{sec:disc} we discuss the 
possible origin of the dust continuum cavity and of 
the whole disk structure. A summary of the main results follows 
in Section~\ref{sec:summary}.

\section{Observations and data reduction}
\label{sec:obs}

\begin{figure*}[!t]
\centering
\includegraphics[angle=0, width=0.8\textwidth]{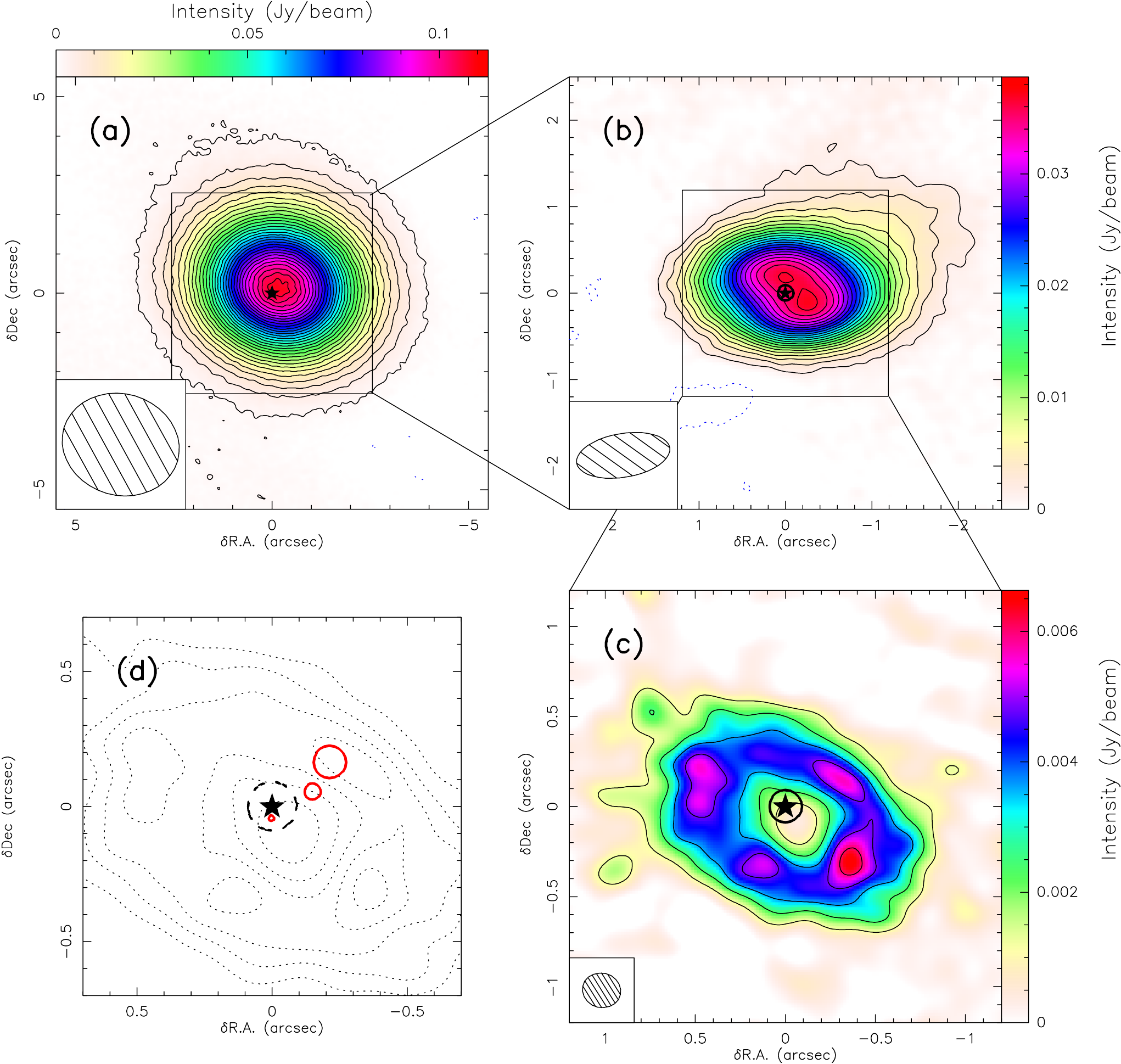}
\caption{\label{fig:obs} Maps of the 1.3 mm dust continuum emission observed toward LkCa~15 
employing three different weighting schemes of the {\it uv} data. {\it (a)} Natural weighting 
was adopted to achieve a FWHM beam size of 3.0\arcsec$\times$2.6\arcsec. 
Intensity contours start at 3$\sigma$ and are spaced by 6$\sigma$, where the $1\sigma$ noise level 
is 0.58 mJy/beam.  {\it (b)} Robust weighting was adopted to achieve a 
resolution of 1.1\arcsec$\times$0.5\arcsec. Contours are plotted every 3$\sigma$, where the 1$\sigma$ 
noise level if 0.53 mJy/beam. {\it (c)} Uniform weighting was adopted to achieve a FWHM beam 
size of 0.21\arcsec$\times$0.19\arcsec. Contours are plotted every 3$\sigma$, where the 1$\sigma$ noise 
level is 0.43 mJy/beam. {\it (d)} The large, medium, and small red circles show the 
position of the center of symmetry of the dust emission corresponding to panel {\it(a)}, {\it(b)}, and {\it(c)} respectively. 
The symbol $\star$ marks the position of the star and the surrounding circle its uncertainty, calculated 
as discussed in the Appendix. The dotted curves show the intensity contours as in panel~{\it(c)}.}
\end{figure*}

LkCa~15 was observed with CARMA array configurations A, B, C, and E, 
providing projected  baseline lengths between 6 and 1700 m. C-configuration observations were obtained in 
October 2007 and are discussed in \cite{Isella09}. Observations in B-configuration 
were obtained  in December 2008. The receivers were tuned at a local oscillator frequency (LO) 
of 224.75 GHz and the continuum emission was observed using three 
contiguous bands of 468 MHz each in both correlator side bands, resulting in a total bandwidth of about 3 GHz. 
Flux and pass band calibration were derived by observing Uranus and 3C84 respectively; 0510+180 was observed 
every 15 minutes to derive the atmospheric and instrumental phase corrections.
A- and E-configuration observations were carried out in November 2010 and January 2011 respectively, using the 
upgraded CARMA correlator which provides a maximum continuum bandwidth of about 8 GHz. 
The receivers were tuned to the same LO frequency used for C and B-configuration 
observations. The upper and lower correlator side bands were configured with eight partially 
overlapping bands of 494 MHz each in order to cover the full bandwidth. 
The band pass shape was calibrated by observing 3C84 and 3C454.3; flux calibration was set by observing 
Uranus and cross checked using almost simultaneous observations of 3C84 and 3C454.3 obtained with the 
SMA at the frequency of 225 GHz. The estimated uncertainty in the absolute flux calibration is 10\%.
Atmospheric and instrumental effects were corrected by observing the nearby 
quasar 0510+180 every 6-10 minutes, depending on the array configuration. 

The data taken in the four different array configurations were independently calibrated using the MIRIAD 
software package. The dataset were shifted both in right ascension and declination to compensate for the stellar 
proper motion (see Appendix) and combined to increase the image sensitivity and dynamic range. 
Images were then obtained through standard inversion of the {\it uv} data 
using different weighting functions to emphasize the morphology of the dust emission on both large 
and small angular scales. The images were deconvolved using the CLEAN algorithm.


\section{Morphology of the continuum emission}
\label{sec:morph}    

The high dynamic range obtained by combining the compact and extended array configurations
enables CARMA to reveal the structure of the dust emission on both small and large angular scales. 
Figure~\ref{fig:obs} shows the map obtained by adopting three different weighting functions to 
achieve an angular resolution of (a) 3.0\arcsec$\times$2.6\arcsec\, (b) 1.1\arcsec$\times$0.5\arcsec\, and 
 (c) 0.21\arcsec$\times$0.19\arcsec. In panel (a), the disk emission appears fairly symmetric and 
centered near the position of the star. The integrated flux measured in a circular aperture 
of 4\arcsec\ in radius, is 128$\pm$5 mJy, and it is consistent with the value 
measured by \cite{Pietu06}. In panel (b), the emission has two peaks separated by about 0.3\arcsec. 
In addition, the low intensity contours show that the disk extends toward the northwest. The integrated flux, 
as measured in a circular aperture of 2\arcsec\ in radius is 127$\pm$15 mJy, where the uncertainty 
accounts only for random noise. Finally, the map shown in panel (c) reveals that the dust emission 
comes from an elliptical ring that extends for about 1.8\arcsec\ along its major axis, and for about 
1.2\arcsec\  in the perpendicular direction. Assuming that the emission arises from a circular disk, 
the aspect ratio of the contour levels leads to a disk inclination of about 50\arcdeg, a position angle 
measured east from north of about 60\arcdeg, and a disk diameter of about 250 AU for the stellar 
distance of 140 pc.  The total integrated flux measured in an elliptical aperture with major axis of 
2.4\arcsec\ and the same aspect ratio of the disk is 117$\pm$10 mJy.  CARMA observations reveal 
for the first time an asymmetric dust distribution in LkCa~15 outer disk, while the ring-like structure 
presented in panel (c) of Figure~\ref{fig:obs} is consistent with earlier lower angular resolution observations 
of LkCa15 \citep[][]{Pietu06,Andrews11a}. 

To investigate the properties of the dust emission we define the center of symmetry $(x_c, y_c)$ of the map
intensity $I(x,y)$ such that $x_c = \Sigma xI / \Sigma I$ and $y_c = \Sigma y I/ \Sigma I$, where the sum 
extends over all positions brighter than 5 times the noise level in the image.
Since $I(x,y)$ depends on the weighting scheme adopted in the image formation process, the position 
of the center of the emission is a function of the angular resolution in the image. 
Panel (d) in Figure~\ref{fig:obs} shows the center of symmetry for the three maps
discussed above.  The stellar position and its uncertainty, as expected from proper motion 
correction (Appendix),  are shown by the star symbol and the surrounding circle respectively. 
We find that the center of symmetry of the highest angular resolution map is consistent with the 
position of the star, and both appear to be shifted  by about 0.1\arcsec\ with respect to the center of 
innermost intensity contours levels. Although the astrometric uncertainty on the position of the 
star does not allow us to firmly conclude on the nature of this shift, we note that a pericenter offset 
of similar magnitude (0.064\arcsec) and direction has been recently derived from infrared 
observations by \cite{Thalmann10}.

At lower angular resolution the center of symmetry moves progressively toward the northwest up 
to a distance of about  0.3\arcsec\ from the star. This shift of the center of symmetry, which results 
from  the asymmetry observed at intermediate angular resolution, suggests that the outermost disk 
regions might have been perturbed. However, about 90\% of the dust emission  
arises from the fairly symmetric ring shown in panel (c). The structure of this ring and the 
significance of the observed asymmetries are discussed in the following two sections.
 
\section{Disk model}
\label{sec:model}

\begin{table*}[!t]
\caption{\label{tab:model} Best-fit parameters for the smooth and truncated models. }
\centering
\begin{tabular}{lllllllll}
\hline
\hline
Model &  Incl.   & PA   &  $r_{in}$ & $r_t$ & $r_{out}$  & $\gamma$ & $\Sigma_t$ & $M_d$  \\
\hline 
Smooth viscous \T \B &    50.5$^{+0.9}_{-0.9}$ & 64.4$^{+1.5}_{-1.5}$ & [0.2]  & 60.2$^{+0.6}_{-0.8}$ & [160] & -2.15$^{+0.08}_{-0.10}$ & 10.37$^{+0.20}_{-0.24}$  &  [0.06]  \\
Truncated viscous \T \B & 52$^{+1.0}_{-1.2}$ & 67$^{+1.6}_{-1.6}$ & 52$^{+1.3}_{-2.0}$ & 20$^{+1.2}_{-1.0}$ & [190] & [1] & 270$^{+20}_{-15}$ &  [0.1] \\
\hline
\hline
Model &  Incl.   & PA   &  $r_{in}$ & $r_0$ & $r_{out}$  & $p$ & $\Sigma_0$ & $M_d$ \\
\hline
Power-law\T \B &   50.7$^{+1.2}_{-1.0}$ &  63.5$^{+1.8}_{-1.2}$ &  42.5$^{+2.4}_{-1.2}$  &  [60]  & 119.7$^{+1.4}_{-2}$  & 0.72$^{+0.14}_{-0.28}$ &  9.92$^{+1.1}_{-0.64}$  & [0.05]  \\
\hline
\end{tabular}
\tablecomments{The numbers in square
brackets are either fixed values (i.e., the normalization radius $r_0$) or are the result of the model fitting. 
The inclination and position angle are expressed in degrees, the radii $r_{in}$, $r_{out}$, $r_t$, and $r_0$ 
in AU, the disk surface densities $\Sigma_t$ and $\Sigma_0$ in g cm$^{-2}$, and the disk mass $M_d$ in solar masses.}
\end{table*}

The disk properties were inferred using two prescriptions for the disk surface density.

The first is a classical power law $\Sigma(r) = \Sigma_0  (r_0/r)^{p}$ characterized by four free 
parameters: the inner radius $r_{in}$, the outer radius $r_{out}$,  the normalization value $\Sigma_0$, and the 
slope $p$. This parameterization provides the simplest form for the disk surface density that mimics the sharp 
inner disk truncation caused by disk photo-evaporation or by dynamical interaction with a giant planet, 
but has the disadvantage of introducing an unphysical sharp edge of $\Sigma(r)$ at the disk outer radius.  

To obtain a smoother form of $\Sigma(r)$ we adopt a second parameterization given by the similarity solution 
for the disk surface density of a viscous keplerian disk \citep[see][]{Lynden74,Hartmann98}, 
adopting the form presented in \cite{Isella09},
\begin{equation}
\label{eq:sigma_used}
\Sigma(r) = \Sigma_t \left( \frac{r_t}{r} \right)^{\gamma} \times  
\exp{ \left\{ - \frac{1}{2(2-\gamma)} 
\left[ \left( \frac{r}{r_t} \right)^{(2-\gamma)} -1 \right] \right\} }.
\end{equation}
This functional form has  five free parameters, i.e., the disk inner and 
outer radius ($r_{in}$ and $r_{out}$), the characteristic radius  $r_t$, the normalization value $\Sigma_t = \Sigma(r_t)$, 
and the slope $\gamma$, which is related to the radial viscosity profile through the equation $\nu(r) \propto r^{-\gamma}$.

Since the limited angular resolution and sensitivity of existing observations does not generally 
allow to fit for all five parameters, previous studies of the dust distribution in transitional disks based on 
this form of the surface density  adopted two different approaches. In \cite{Isella09,Isella10a,Isella10b}, we fixed the disk inner radius at 
the dust evaporation radius ($< 0.5$ AU), and compared the model to the observations to constrain the values 
of $\gamma$, $r_t$, and $\Sigma_t$. In this way, we found that the cavities in the dust continuum 
emission observed in LkCa~15, RY~Tau, and MWC~758 disks might be explained with small inner disk radii and 
negative values of  $\gamma$, which leads to a smoothly increasing surface density up to a radius $r_{max} = 
r_t \times (-2\gamma)^{1/(2-\gamma)}$.  A different approach was adopted by \cite{Andrews11a},
who fixed the slope $\gamma = 1$, and derived the disk inner radius $r_{in}$,  
$r_t$ and and $\Sigma_t$\footnote{ \cite{Andrews11a} adopt a different parameterization of the similarity solution 
for the surface density, which depends on the characteristic radius $r_c = r_t \times [2(2-\gamma)]^{1/(2-\gamma)}$ 
and on the normalization value $\Sigma_c = \Sigma_t \times [2(2-\gamma)]^{\gamma/(\gamma-2)}$. }
 The first modeling procedure, i.e., variable $\gamma$ and fixed $r_{in}$, 
 stems from the idea that transitional disks might be shaped a global physical process, such as 
 the disk viscosity and/or the variation of the dust opacity due to the grain growth. By contrast, 
 the second case assumes that the disk surface density profile follows the $\alpha$-constant prescription of the 
 viscosity \citep{Shakura1973} which leads to $\gamma=1$, and that the cavities observed in the 
 dust emission results from the inner truncation of the disk, perhaps due to planets, 
 which does not effect the density distribution at larger radii. In order to investigate the nature 
 of LkCa~15 disk we here adopt both 
 approaches which we will be referred to as {\it smooth} and {\it truncated} viscous models.


The radiation transfer equation within the disk is solved by adopting the ``two-layer" approximation 
of \citet{Chiang97}, in which the disk is described through a warm surface layer and a cooler 
midplane interior. Both temperatures are calculated as a function of the orbital radius by iterating on 
the vertical disk structure \citep[see][]{Dullemond01}. 
The disk is in hydrostatic equilibrium between the gas pressure and the stellar gravity, which leads to 
a flared geometry with the opening angle increasing with the distance from the central star. Because a flared 
disk geometry tends to over predict the far-IR disk emission in LkCa 15, we allow the disk scale height 
$H_p$ to be lower than the self-consistent solution $\tilde{H}_p$ by introducing a settling parameter $\Psi < 1$, 
so that $H_p=\Psi\tilde{H}_p$. Following \cite{Mulders10} and \cite{Espaillat07}, we chose $\Psi = 0.3$.

The dust opacity is calculated by assuming an interstellar grain composition 
\citep{Pollack94} and a particle size distribution between a minimum ($a_{min}$) 
and a maximum ($a_{max}$) size  according to $n(a) \propto a^{-q}$. The spectral 
slope of the LkCa15 mm-wave disk emission is $\alpha=3.4$, as computed from the 
observed flux density at 1.3 mm (128 mJy; see Section 2) and 7 mm \citep[0.44 mJy; ][]{Rodmann06}.
Assuming that the 
emission at these wavelengths is optically thin and follows the Rayleigh-Jeans approximation, 
this implies a dust opacity slope  $\beta=2-\alpha=1.4$, which is reproduced adopting  
$a_{min} = 0.5$~$\mu$m, $a_{max} =0.5$~mm, and $q=3.5$. 
For sake of simplicity  we assume that the dust opacity is constant throughout the disk. 

From the derived dust density, temperature, and opacity we calculate the 
disk SED and synthetic disk images in the dust continuum at 1.3~mm. We assume 
that the center of the disk corresponds to the center of symmetry of the dust emission 
shown in panel (c) of Figure~\ref{fig:obs}. The synthetic disk images are then Fourier 
transformed and sampled at the positions on the (u,v) plane corresponding 
to the CARMA observations. Best-fit values for the free parameters that define the disk 
surface density, and for the disk inclination and position angle are then derived through 
a $\chi^2$ minimization based on a Markov Chain Monte Carlo
simulation as discussed in \cite{Isella09}.

\section{Results}
\label{sec:res_mod}

\subsection{Constraints on the disk surface density}
\label{sec:res_sigma}

\begin{figure*}[t]
\centering
\includegraphics[angle=0, width=0.9\textwidth]{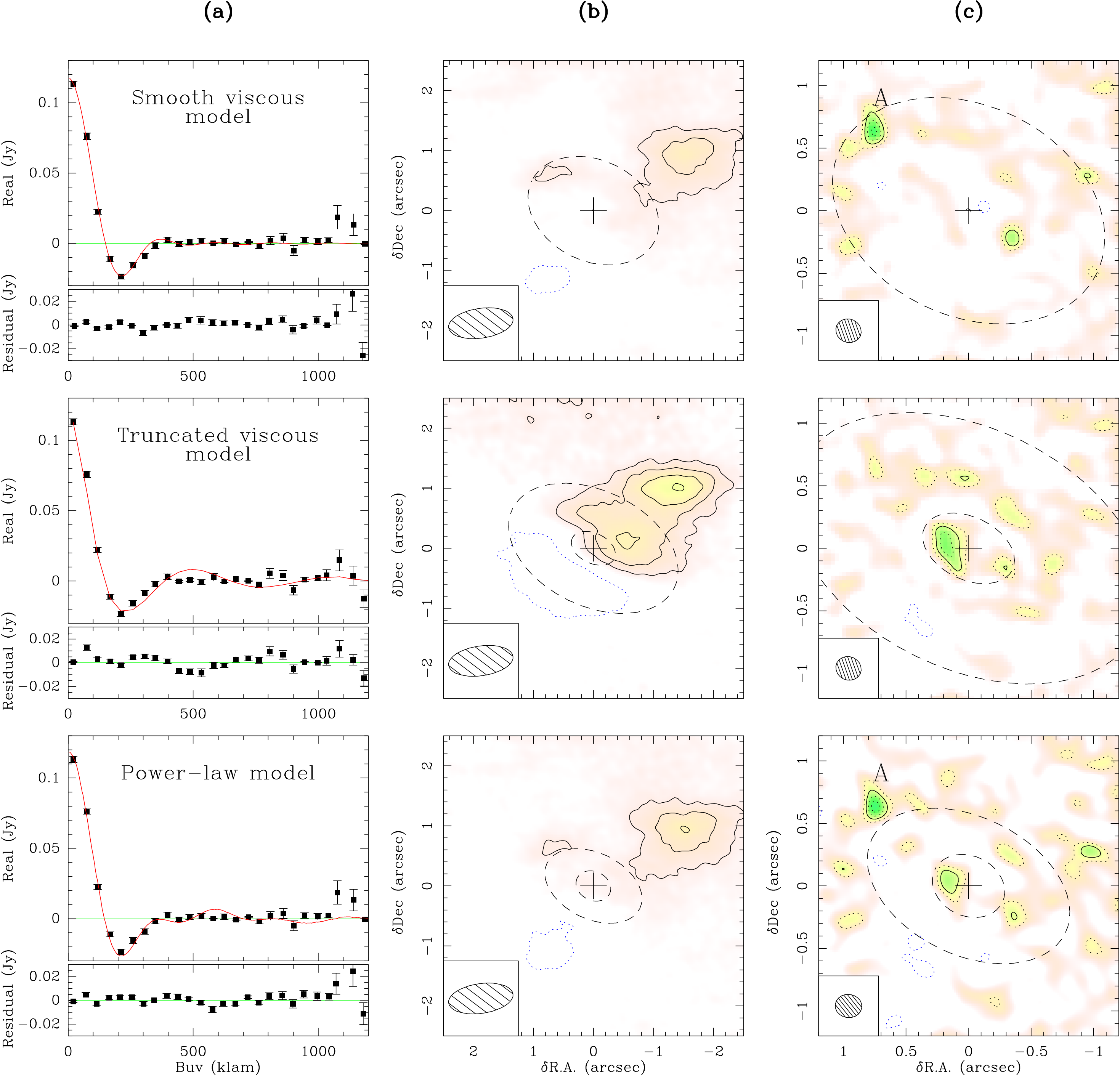}
\caption{\label{fig:mod} Comparison between the observations and the best-fit 
models calculated assuming a smooth viscous (top panels), a truncated viscous (middle panels) 
and a power-law (bottom panels) 
surface density profile. Panels (a) show the real part of the correlated flux observed 
at 1.3 mm (square points with error bars), along with the best-fit model (solid line) and 
the residuals. To account for the disk geometry, the data have been 
deprojected using the values for the inclination and position angle 
listed in Table~\ref{tab:model}. Panels (b) and (c) 
show maps of the residuals obtained with the weighting functions used in panels (b) 
and (c) in Figure~\ref{fig:obs}. The color scale,  the map size and 
noise level are also the same as in Figure~\ref{fig:obs}. 
Solid contours are plotted every 3$\sigma$, while the dotted contours 
in panels (c) stars at  2$\sigma$ and are spaced by 1$\sigma$. The dashes ellipses show 
the inner and outer disk radius corresponding of the best-fit models  listed in Table~\ref{tab:model}.
}
\end{figure*}

The best-fit parameters for both the smooth and the truncated disk models 
are listed in Table~\ref{tab:model}.  The quoted uncertainties correspond to 
the 68.3\% confidence level  and were calculated by integrating the marginalized probability distribution resulting 
from the Monte Carlo fitting procedure \citep[see Appendix A of ][]{Isella09}.
In addition to the free model parameters, we list within the square brackets some relevant 
values for the disk structure, such as the inner and outer disk radius for the 
smooth viscous model, the outer radius and $\gamma$ for the viscous truncated 
disk model,  and the normalization radius for the power-law model. 
Finally, the last column of the table shows the total disk mass. 
In addition to the statistical errors,  the surface density $\Sigma_t$ and the total disk 
mass M$_d$ are affected by systematic errors due to the 
flux calibration and to the uncertainty on the dust opacity. 
We estimate a 10\% uncertainty from flux calibration based upon CARMA 
flux calibration monitor programs. The uncertainty
on the dust opacity is more difficult to quantify and might vary between 5\% and 20\% once 
the slope of the millimeter SED is used to constrain the grain size distribution \citep[see the discussion in ][]{Isella09,Isella10a}.

\begin{figure}[!t]
\centering
\includegraphics[angle=0, width=\columnwidth]{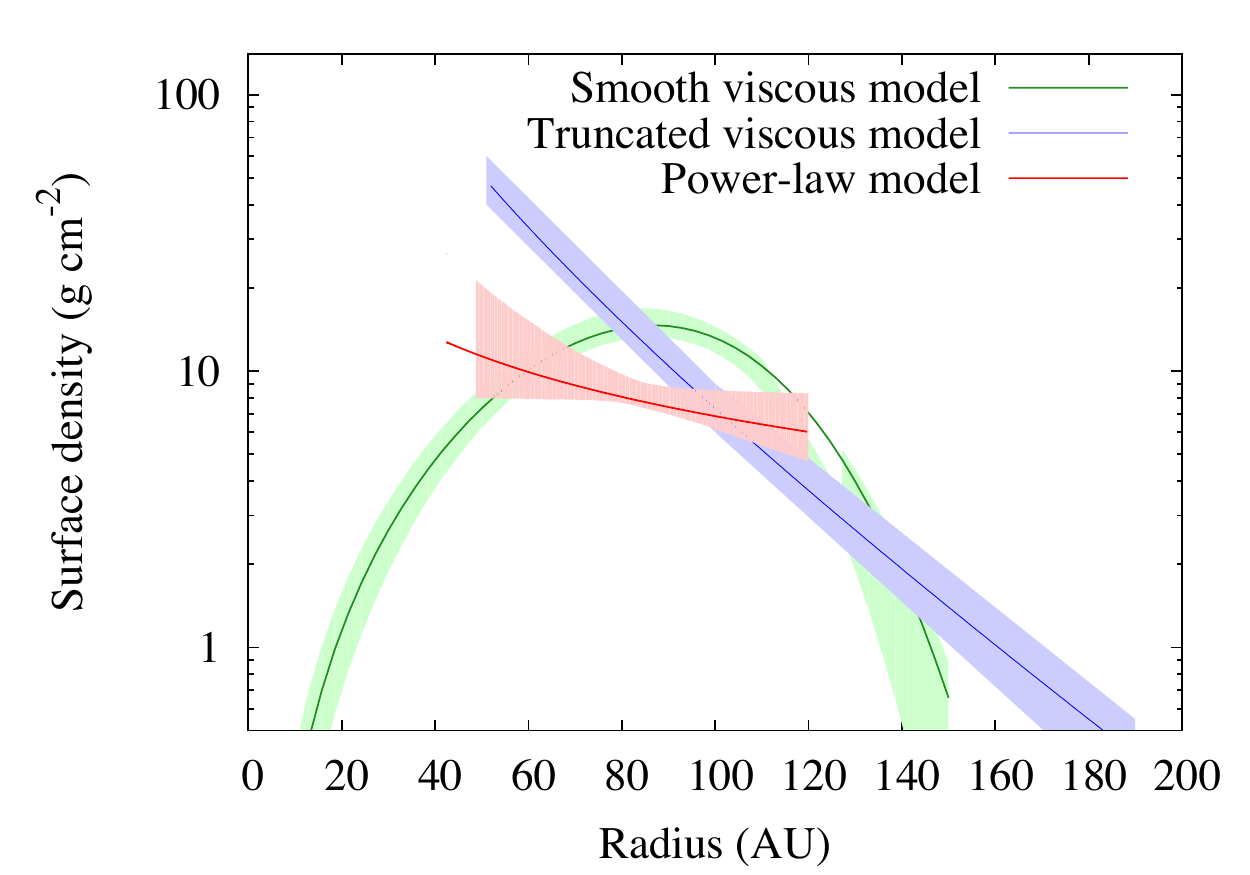}
\caption{\label{fig:sigma}  Surface density profiles corresponding to the 
smooth and truncated viscous best-fit models (green and blue respectively), and to the 
power-law model (red). The colored regions represent the 3$\sigma$ 
uncertainty interval.}
\end{figure}

Figure~\ref{fig:mod} presents the comparison between the best-fit models and 
observations.  Panels (a) show the real part of the observed correlated 
flux (points with error bars), the visibility profile of the best-fit model (solid line) 
and the residuals obtained by subtracting the model visibility to the 
observations. Panels (b) and (c) show the maps of the residuals 
obtained adopting the same weighting functions and color scale 
of panels (b) and (c) of Figure~\ref{fig:obs} respectively.
Finally, the surface density profiles corresponding to the  best-fit models are 
shown in Figure~\ref{fig:sigma}, along with the corresponding uncertainties.  

We find that the smooth viscous and the power-law models provide 
acceptable fits to the observations. The first model is characterized by $\gamma = -2.15$,
 which is more negative, although consistent within the uncertainties, with the $\gamma$ value 
 derived in 2009 by fitting only the smaller spatial frequencies.
 Both models suggest a disk inclination of 51\arcdeg$\pm$1\arcdeg\ and a position angle of 64\arcdeg$\pm$2\arcdeg, that
agrees with the values derived by \cite{Andrews11a}.
By contrast, the truncated viscous model characterized by $\gamma=1$ does a 
worse job in reproducing the structure of the dust emission and leads to significative 
residuals both inside the continuum cavity and along the dusty ring (second row of Figure~\ref{fig:mod}).
The comparison between the best-fit surface densities (Figure~\ref{fig:sigma}) shows that 
the first two models are characterized by rather flat density profiles between about 
50 and 120 AU, while the truncated viscous model produces a much 
steeper decrease in the surface density in this radial range. This result therefore
implies that the surface density in LkCa~15 disk 
is  inconsistent with the $\alpha$-constant prescription at least in the radial 
range probed by our observations. 

\begin{figure}[!t]
\centering
\includegraphics[angle=0, width=\columnwidth]{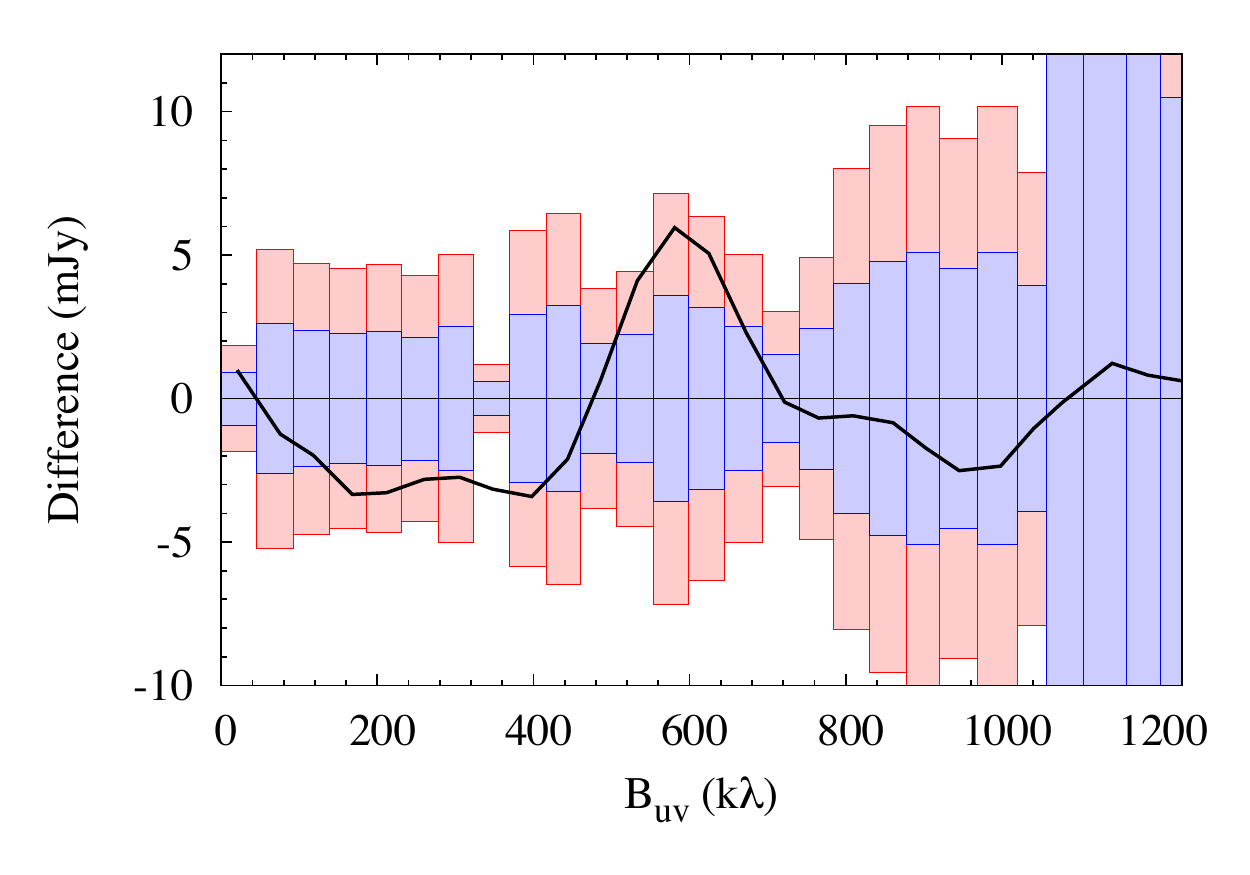}
\caption{\label{fig:comparison} Difference between the 1.3~mm correlated fluxes 
(solid curve) of the best-fit power-law and smooth viscous models. The blue and red triangles 
show the 1$\sigma$ and 2$\sigma$ noise level of the CARMA observations. The figure
shows that the sensitivity of CARMA observations is not sufficient to distinguish between 
the two surface density parameterizations. 
}
\end{figure}

The power-law and the smooth viscous models reproduce equally well 
the ring observed in the dust emission. To understand why our observations cannot 
disentangle these two models, we show in Figure~\ref{fig:comparison} the difference 
between the correlated flux of the power-law and the 
smooth viscous model as a function of the spatial frequency (solid curve), along with 
the 1$\sigma$ and 2$\sigma$ noise level provided by our observations (blue and red 
boxes respectively). The maximum difference between the two models is about 
6 mJy at 580 k$\lambda$ and is within the 2$\sigma$ noise level.
The figure also shows that our ability to disentangle different surface density profiles 
is not hampered by the angular resolution (i.e., by the maximum spatial frequency 
achieved by our observations),  but by the noise level on spatial frequencies between 
100~k$\lambda$ and 700~k$\lambda$ where the difference between the two models 
is the greatest. This is an important point to consider in planning future observations 
and will be discussed further in  Section~\ref{sec:alma}.

Although we cannot achieve a model-independent constraint of 
the disk surface density,  our observations provide an upper limit of the 
amount of small solid particles present in the LkCa15 inner disk.  We find that 
the upper limit is set by the best-fit smooth viscous  model, since any additional amount 
of dust in the inner disk provides a worse fit to the data. Integrating this
density profile between 3-42 AU, we calculate a dust mass of 7 Earth masses (M$_\oplus$), or 
about 2 M$_J$ of gas assuming a standard gas-to-dust ratio of 100. 
Additional constraints of the gas density come from spatially resolved 
observations of the $^{12}$CO and $^{13}$CO (2-1) line emission \citep{Pietu07}. 
These observations are characterized by an angular resolution between 
1\arcsec-2\arcsec, and were compared to power-law disk models to derive the 
gas density and temperature radial profile. \cite{Pietu07} find  that CO might
be present down to a radius of 10 AU, and that a sharp truncation of the gaseous disk
at 50 AU is excluded at the 7$\sigma$ level. These observations also suggest the presence 
of  about 2 M$_J$ of gas within 42 AU from the central star. The amount 
of dust within 42 AU from the central star is also constrained by the infrared SED 
modeling, which is discussed in Section~\ref{sec:sed}.

Finally, we confirm the discrepancy found by \cite{Pietu07} between  the disk outer 
radius inferred from the dust continuum emission, i.e., about  160 AU,  and that derived from 
the observations of the molecular line emission, which, depending of the 
observed molecule, varies between 550 AU ($^{13}$CO) and 900 AU ($^{12}$CO).  
Gaseous disks are found systematically larger than dusty disks in almost 
all the objects observed at high angular resolution \citep{Isella07,Pietu05,Pietu07}.
In some cases this discrepancy can be reconciled by adopting an exponentially tapered 
surface density profile, as the one resulting from the disk viscous evolution \citep{Hughes08,Isella09}.
This model predicts that dust is present up to the outer radius inferred from the gas but the 
millimeter-wave dust emission coming from the outer disk is too low to be detected by existing 
observations. In the case of LkCa~15,  we have used the gas emission model 
discussed in  \cite{Isella07} to calculate the $^{12}$CO J=2-1 emission corresponding to 
the smooth and truncated viscous best-fit models for the dust continuum emission. 
The $^{12}$CO surface density is described by Equation~\ref{eq:sigma_used}, with $\gamma$ and $r_t $ 
as in Table 1, and assuming a CO/H2 abundance of $10^{-4}$, as measured in the molecular clouds. 
By assuming that the $^{12}$CO molecules are in thermal equilibrium with the dust, 
we find that  the $^{12}$CO emission should be confined within 300 AU and 550 AU from the central star
for the smooth and truncated viscous model respectively, despite the fact that the $^{12}$CO density extends in practice to infinity. 
The fact that $^{12}$CO is observed out to a radius of 900 AU suggests therefore 
that the exponential taper that characterizes the viscous surface density is not 
a good explanation for the different radial extent of  the gas and dust emission 
in LkCa~15. Other possible explanations are discussed  in Section~\ref{sec:dis_radius}.


\subsection{Sub structures and asymmetries}
\label{sec:residuals}
Figure~\ref{fig:mod} shows the map of the residuals obtained by subtracting the 
best-fit models from the observations in the Fourier domain, and by adopting different 
weighting functions in the image reconstruction process. Panels (b) and (c) 
adopt the same weighting functions, use the same color scale, and have the 
same size of the maps shown in the panel (b) and (c) of Figure~\ref{fig:obs}.

Panel (b) clearly shows the asymmetric emission in the northwest side of the disk 
discussed in Section~\ref{sec:morph}. The integrated flux is about 8 mJy, which 
corresponds to a dust mass between 5 and 10 M$_J$ for a dust temperature between 
20-40 K.  

As for the residuals along the dusty ring and inside the cavity, we discuss only the 
case of the smooth viscous and power-law model, since we have concluded in the 
previous section that the truncated viscous model does not provide as good of a fit to the observations.
For these two models, the brightest residual in panels (c) is identified with the letter A and 
has peak flux of 2.4 mJy. It is located along the major axis of the disk at the physical distance 
of 140 AU from the center, and therefore very close, or just outside, the outer disk radius inferred from 
our disk modeling, which is represented by the large dashed ellipses in Figure~\ref{fig:mod}. 
At this distance from the central star, the dust is expected to have a temperature of about 
40 K, which leads to a mass of dust of about 1.3 M$_J$. 

\begin{figure}[!t]
\centering
\includegraphics[angle=0, width=\columnwidth]{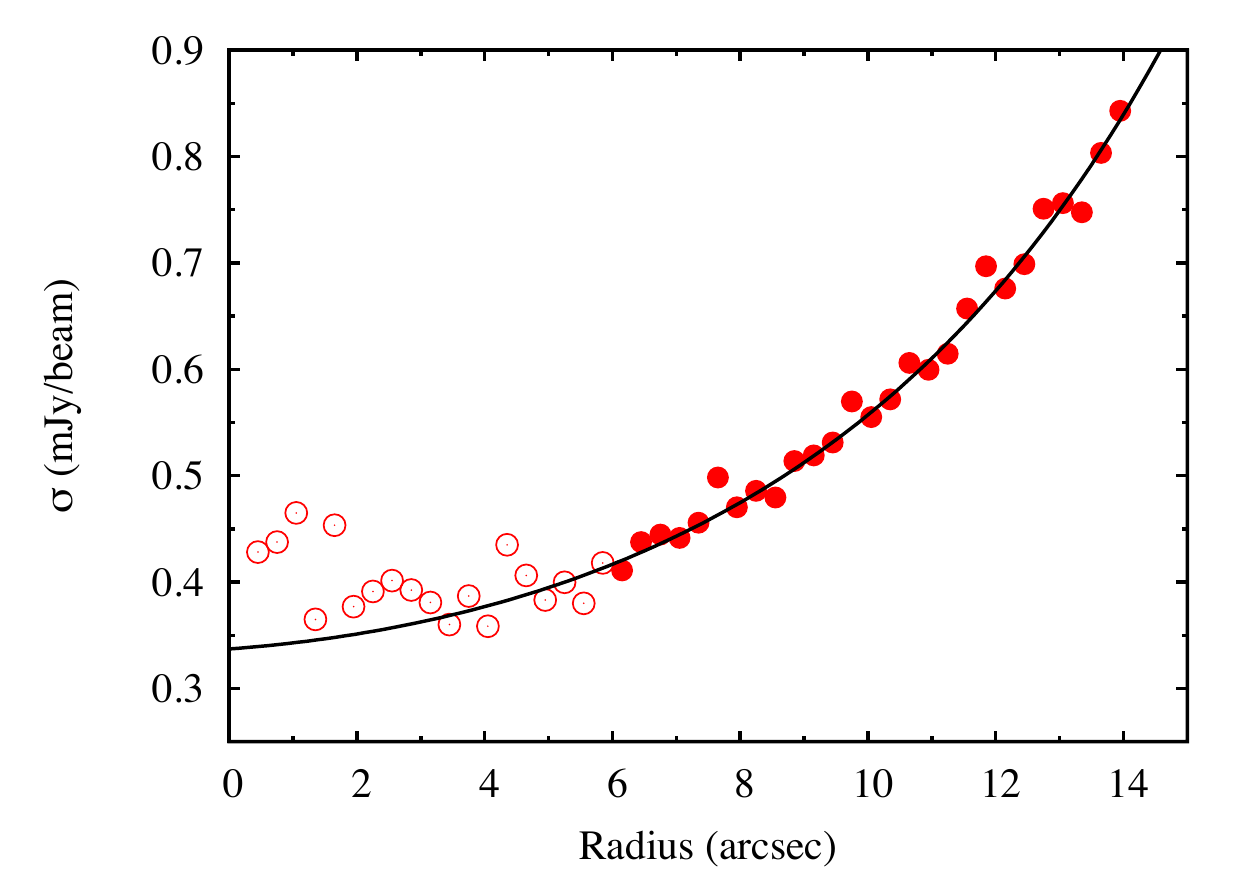}
\caption{\label{fig:noise}  Radial variation of the noise in the map presented 
in panel (c) of Figure~\ref{fig:mod}. The circles show the standard 
deviation ($\sigma$) of the intensity measured in concentric 0.2\arcsec 
wide annuli. The filled circles correspond to annuli in which the intensity 
distribution is consistent with a standard normal distribution, as expected 
if the emission is due to gaussian noise. The open circles indicate the radii 
where the intensity distribution is dominated by true emission.
The solid curve represents the theoretical noise as computed from the weights 
on the visibility data.}
\end{figure}

The significance of this residual depends critically on knowing the value of the noise 
in our map and how it varies with the distance from the center due to the antenna primary 
beam attenuation. A theoretical value for the noise can be obtained from the 
weights on each visibility point, which are measured at the time of the observations and 
depends on the effective area of the telescope and the system temperature 
\citep[see, e.g, ][]{Thompson91}. However, if the observed source is bright, 
this estimate does not take into account possible effects due to the image deconvolution, 
such as the presence of residual side lobes in the final map, and dynamic range 
limitations due to the atmospheric decorrelation (i.e., seeing) and calibration errors.

We estimated the noise in the images as follow. We divide the intensity map in concentric 
annuli characterized by a width of 0.2\arcsec. For each annulus we calculate the mean 
intensity and the standard deviation $\sigma$. If the emission is due only to gaussian noise, 
then the mean is 0 and the standard deviation of the intensity is equal to the noise. For each 
annulus, we check that the distribution is gaussian by creating a histogram for the intensity 
and fitting a normal distribution. Otherwise, if the annulus contains true emission, the intensity 
distribution will deviate from the normal profile. In this case $\sigma$ will give an upper estimate 
of the noise. 

Figure~\ref{fig:noise} shows the radial variation of the 
standard deviation of the intensity, $\sigma$.  At radii larger 
than 6\arcsec\ (filled circles), the intensity distribution is well described by a 
standard normal function; $\sigma$ increases with the radius as the inverse of the 
antenna primary beam profile and it is equal to the theoretical 
noise indicated by the solid curve.  Within 6\arcsec, the intensity is a combination of 
gaussian noise and true emission and its distribution deviates from 
the standard normal profile. In this region, the noise in the map 
may assume values between the theoretical noise and $\sigma$.
This implies a noise upper limit of 0.47 mJy/beam at the position 
of the residual A (i.e., 1\arcsec), which leads to a signal-to-noise 
ratio of 5.1, given the measured peak flux of 2.4 mJy. 

The significance level of residual A can be estimated by multiplying the normal 
probability corresponding to a 5.1$\sigma$ residual ($3.4 \times 10^{-7}$) by the number 
of independent pixels in our map. If each resolution element (0.21\arcsec$\times$0.19\arcsec) 
is accounted as independent pixel, then the number of independent pixels within the 1.3 mm 
CARMA field of view (27\arcsec) is about 18000, and the probability that the residual might be 
due to random noise is smaller than 0.6\%, or about 2.8$\sigma$. This result suggests that the 
residual might trace a real compact over-density in the dust distribution with respect to the best-fit 
surface density profile, but more sensitive observations are required to investigate the nature of this 
emission.  

All the other residuals in panel (c) of Figure~\ref{fig:mod} 
have flux within $\pm$4.5 times the noise level. The probability that any one of these 
residuals could originate from random noise is greater than 10\%. 

 \subsection{Understanding the SED}
\label{sec:sed}

\begin{figure}[t]
\centering
\includegraphics[angle=0, width=\columnwidth]{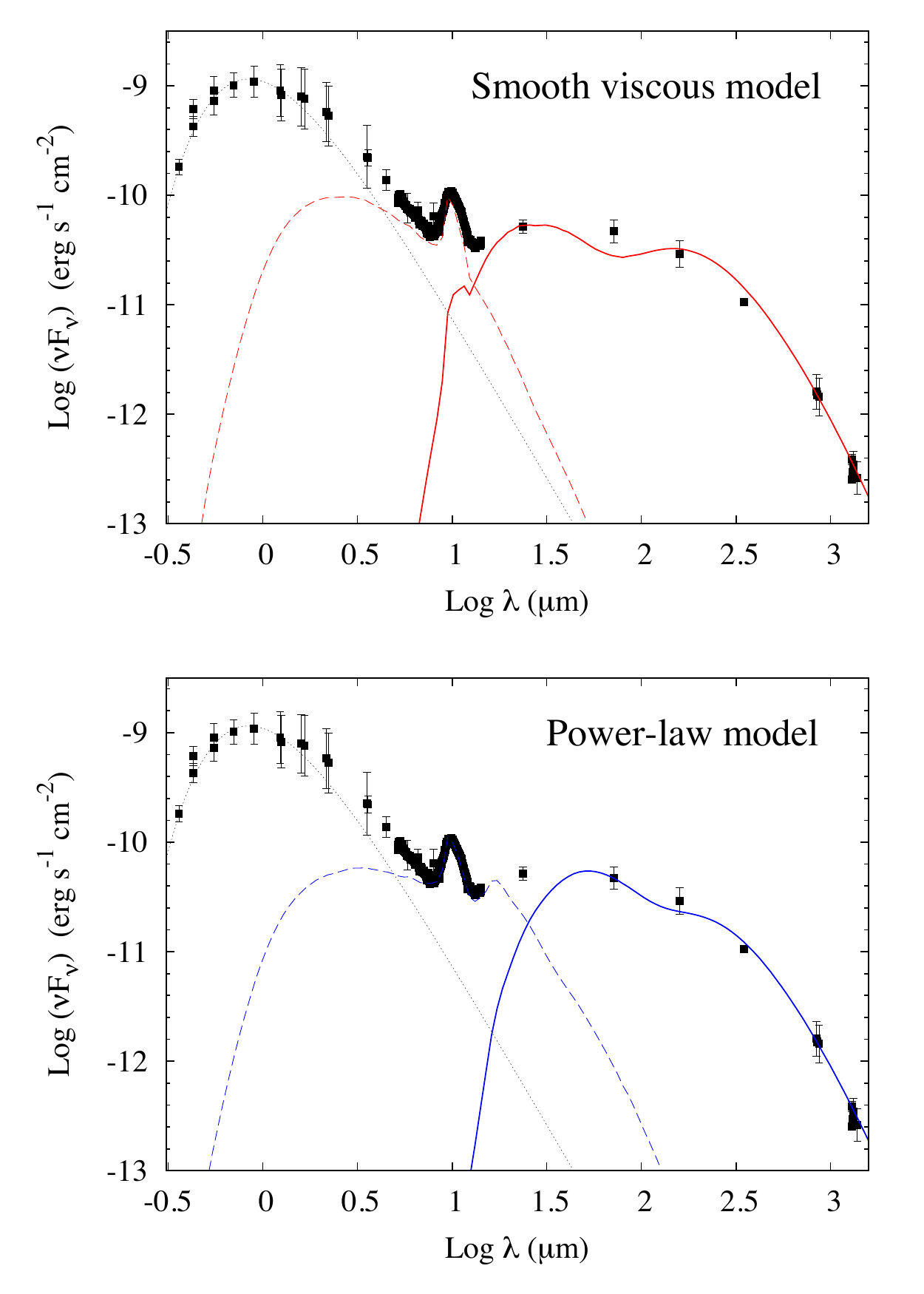}
\caption{\label{fig:sed} SED of LkCa~15. Filled squares present photometric data from 
the literature \citep{Kenyon95,Hartmann05,Espaillat07,Rebull10,Kitamura02,Dutrey96,
Andrews05,Andrews07,Oberg10} and from our new CARMA observations. Infrared 
photometry at 9~$\mu$m, 19.6~$\mu$m, and 90~$\mu$m is from the Akari archive, and 
the infrared spectrum from 5~$\mu$m to 14~$\mu$m is from archival Spitzer IRS 
observations. The solid curves show the SED of the best-fit models for the smooth (top panel)
and the power-law (bottom panel) surface density parameterization. The dashed curves indicate 
the SED required to reproduce the observed near- and mid-infrared excess. As discussed in 
the text, this additional component can arise either from a narrow ring of optically thick dust, 
or from a  more extended optically thin region.}
\end{figure}

Figure~\ref{fig:sed} compares the observed SED for LkCa~15 with the SED inferred from the best-fit
models to the millimeter data. Both the smooth viscous (upper panel) and 
the power-law (lower panel) provide a good agreement with the observations at
millimeter wavelengths. At infrared wavelengths, the model SED strongly
depends on the disk surface density profile. In the case of the smooth viscous 
model, the disk is optically thick to the stellar radiation starting from 3 AU, 
and the agreement with the observation extends to about 20~$\mu$m. At
shorter wavelengths the model SED underestimates the observed emission 
from the disk. This is
due to the fact that the surface density within a few AU from the central star is lower than 
$10^{-6}$ g cm$^{-2}$ and the near-infrared disk emission is negligible with respect to the 
stellar photosphere. In the case of the power-law model, the predicted SED agrees with the 
observation only at wavelengths longer than 50~$\mu$m due to the cut-off of the dust 
density at 42.5 AU. Independently on the adopted model, the near-infrared emission observed 
toward LkCa~15 requires the presence of material with temperature  between 1000-2000 K in 
excess of what predicted by our models.  

For a power-law surface density profile truncated 
at 42.5 AU,  the near infrared emission can be explained by two 
different models, as discussed in \cite{Mulders10}. In the first case, the emission 
comes from a small optically thick disk that extends from the dust sublimation 
radius ($\sim$0.1 AU) up to about 1 AU \citep[see also][]{Espaillat07,Espaillat08}. 
For typical infrared dust opacities, this model requires a minimum dust mass 
of $3 \times 10^{-5}$ M$_\oplus$. Alternatively, the LkCa15's infrared SED can 
be fitted with an optically thin region that extends from 0.1 AU to 5 AU.  
In this case, the mass of dust would be $3 \times 10^{-6}$ M$_{\oplus}$.
The optically thin region and the optically thick disk would produce a 1.3 mm 
flux below  $10^{-6}$ mJy and would therefore have a negligible effect 
on the overall disk millimeter emission. The same two models for the near-infrared 
emission can be applied to the case of the smooth viscous surface density profile. However 
in this case the outer disk is brighter in the mid-infrared than in the power-law case, 
and therefore the region emitting the infrared emission has to be smaller.

\section{Discussion}
\label{sec:disc}

\subsection{On the origin of the continuum cavity}
\label{sec:disc_inner}

The results presented so far suggest at least two possible scenarios
for the radial distribution of the circumstellar material around LkCa~15.

In the first case, most of the circumstellar dust is confined in a sharply truncated 
ring extending between 42-120 AU. Since no discontinuity is observed in the gas 
density at the edges of this dusty ring, this scenario requires a drop 
of the dust-to-gas ratio, or of the dust opacity, outside the ring. As discussed in 
Section~\ref{sec:intro}, a sharp internal disk truncation can be the result of 
dynamical interactions with one, or more, giant planets \citep[see, e.g., ][]{Bryden99} or 
be due to the disk photoevaporation caused by the stellar radiation field \citep[see, e.g., ][]{Gorti09}. 
This latter model suffer however two severe problems. First, it predicts a truncation 
in both the gaseous and dusty disk, which, as discussed in Section~\ref{sec:res_sigma}, it is 
not consistent with spatially resolved observations of the mm-wave CO emission. Second, 
it predicts that the accretion of material on the central star stops as soon as the cavity is opened, 
while the mass accretion rate measured toward LkCa~15 is  about $\dot{M} = 3 \times 10^{-9}$ 
M$_\sun$ yr$^{-1}$ \citep{Hartmann98}. 

Whether  large ($>$ 20 AU) dust depleted cavities can be produced by the dynamical 
interaction with giant planets is still subject of debate. \cite{Zhu11} showed 
 that for large values of the viscosity parameter ($\alpha > 0.01$), 
a system with 3 or 4 Jupiter mass planets orbiting between 5 and 20 AU can lead to a 
largely depleted gap in the disk surface density. However, the mass 
accretion rate on the central star is predicted to drop below $5\times10^{-10}$ 
M$_\sun$ yr$^{-1}$ in less that 1 Myr after the formation of the planets. 
Furthermore, a discontinuity in the gaseous disk surface density will not be consistent with the CO
observations. Alternatively, a lower viscosity ($\alpha \leq 5\times10^{-3}$) would 
lead to larger values of $\dot{M}$ and less depleted cavities. While this latter model might 
be consistent with the observation of the CO emission, it fails to explain the lack of mid-infrared 
dust continuum emission observed
toward LkCa~15. To reconcile the measured values of $\dot{M}$ with the infrared SED 
in the framework of planet-disk interaction,  \cite{Zhu11} argue that low values 
of $\alpha$ have to be coupled with a decrease by at least one order of magnitude of the 
near- and mid-infrared dust optical depth in the region partially depleted by the planets.
A drop of this magnitude in the dust opacity can be achieved through the coagulation of 
small, micron sized grain into larger bodies. In the particular case of LkCa~15, we find that 
the dust opacity at 10~$\mu$m will decrease from  about 3 cm$^2$ g$^{-1}$ to 0.2 
cm$^{2}$ g$^{-1}$ by increasing the maximum grain size from the adopted value 
of 0.5 mm (Section~\ref{sec:model}) to 10 cm. It is therefore possible that the actual structure 
of the LkCa~15 inner disk is the result of an earlier phase characterized by the formation
of large pebbles and boulders in the innermost disk regions \citep[see, e.g.,][]{Brauer08}, 
followed by the formation of giant planets.  If that happened in a low viscosity 
environment, the gas density within 40 AU would be high enough to explain both 
the measured mass accretion rate and the CO observations, and perhaps to allow 
the planetary cores to accrete the gas required to reach the Jupiter mass.

\begin{figure}[!t]
\centering
\includegraphics[angle=0, width=\columnwidth]{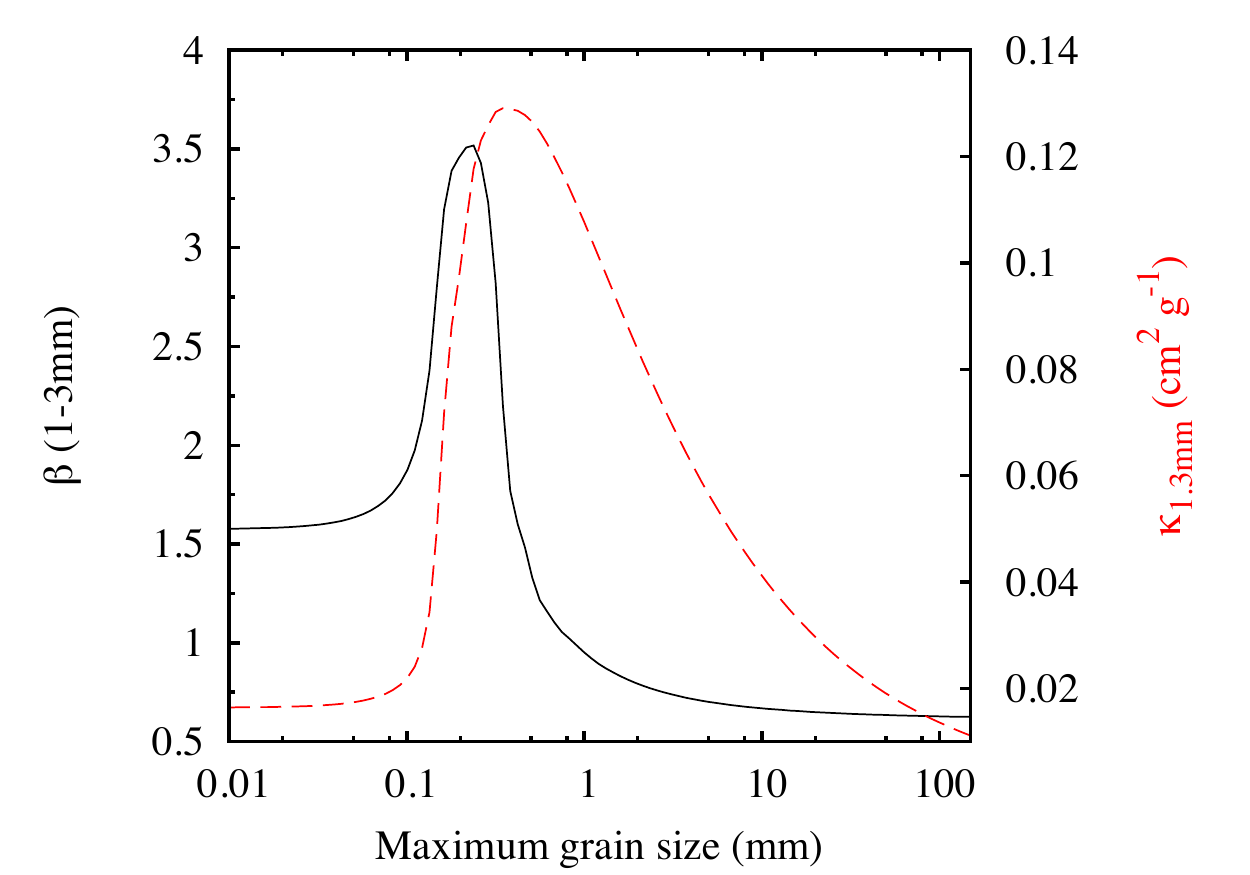}
\caption{\label{fig:beta} Dependence of the slope of the dust opacity $\beta$ measured 
between 1.3 and 3~mm (solid curve) and of the dust opacity $\kappa$ at 1.3 mm 
(dashed curve) on the maximum grain size, for a grain size distribution $n(a) \propto a^{-q}$, 
where $q=3.5$ and the minimum grain size is $5 \times 10^{-6}$ mm.}
\end{figure}

This scenario is compatible with the recent discovery of a possible proto-planet orbiting 
at about 16 AU from the central star discovered through aperture masking observations 
in the $K'$ ($\sim$2.1~$\mu$m) and $L'$ ($\sim$~3.8$\mu$m) bands  \citep{Kraus11}. 
The $K'$ magnitude and $K'-L'$ color of the planet are consistent with a temperature of 
1500~K and a mass of 6~M$_J$, or smaller. In addition the $L'$ observations trace extended 
emission interpreted as evidence of circumplanetary material. However, following theoretical 
models for the planet disk interactions, this 
planet would be able to open only a relatively small gap in the dust distribution between 
12-20 AU.  To explain the cavity radius of 42 AU, at least another planet clearing the disk 
within 12 AU, and other two planets clearing the region between 20 and 42 AU would 
be required \citep[see, e.g.,][]{Dodson11}.

In an alternative scenario, which does not require multiple planets,  the disk surface 
density is described by the smooth viscous model shown in Figure~\ref{fig:sigma}.
In this case, the gas density decreases within 80 AU but it remains high enough
to explain the small inner disk radius derived from the CO observations.
This situation is similar to that proposed for MWC~758, where the $^{12}$CO emission is 
centrally peaked while the dust emission traces a partially dust 
depleted disk within 70 AU from the central star \citep{Isella10b}.  It is important 
to note that the smooth best-fit model was derived in the assumption 
of constant dust opacity throughout the disk. By relaxing this assumption, we can interpret 
our results in terms of radial variation of the dust opacity, due, for example, to grain growth.  
Radial variations in the grain size distribution are indeed predicted by theoretical models 
for the dust evolution \citep{Brauer08} and can be observed by measuring radial variations 
of the mm-wave opacity slope $\beta$ through spatially resolved multi-wavelengths 
observations of the dust emission \citep{Isella10a, Guilloteau11}. In the case of LkCa~15, 
the comparison of 1.3 and 3 mm observations suggests that $\beta$ increases 
from 0.7-1.0 to 1.5-1.7 between 30 AU and 150 AU \citep{Guilloteau11}. As shown in 
Figure~\ref{fig:beta}, this variation would correspond to a decrease of the maximum grain 
size from about 8-30 mm to 0.3-3~mm, or to an increase of about a factor of 2 in the dust 
opacity at 1.3 mm. The observed variation in the grain size might therefore explain 
the observed cavity in the 1.3~mm dust emission without invoking multiple giant 
planets, but more sensitive observations at longer wavelengths are required to investigate 
the presence of large grains in the 1.3 mm continuum cavity.

\subsection{Investigating the origin of transition disks with ALMA}
\label{sec:alma}

\begin{figure}[!t]
\centering
\includegraphics[angle=270, width=\columnwidth]{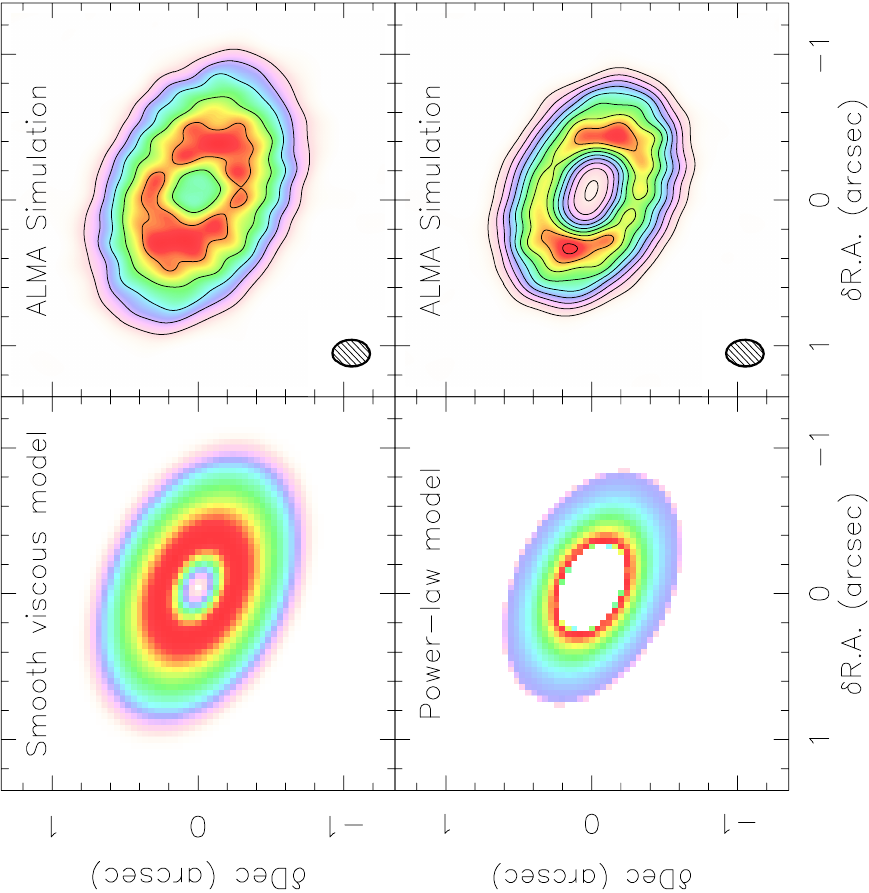}
\caption{\label{fig:alma} Simulated ALMA observations of 0.44 mm continuum emission 
corresponding to the smooth viscous (top panels) and power-law (bottom panels) best-fit models 
discussed in Section~\ref{sec:res_sigma}. The panels on the left show the model 
while the simulated observations are shown on the right column. 
The simulations assume a 30 min on source integration and the default weather conditions 
for Band 9. ALMA will determine whether the inner 
edge of the disk is sharply truncated (possibly due to dynamical interactions 
with a planet), or varies smoothly with radius (possibly due to grain growth in the inner disk).}
\end{figure}

\begin{figure}[!t]
\centering
\includegraphics[angle=0, width=\columnwidth]{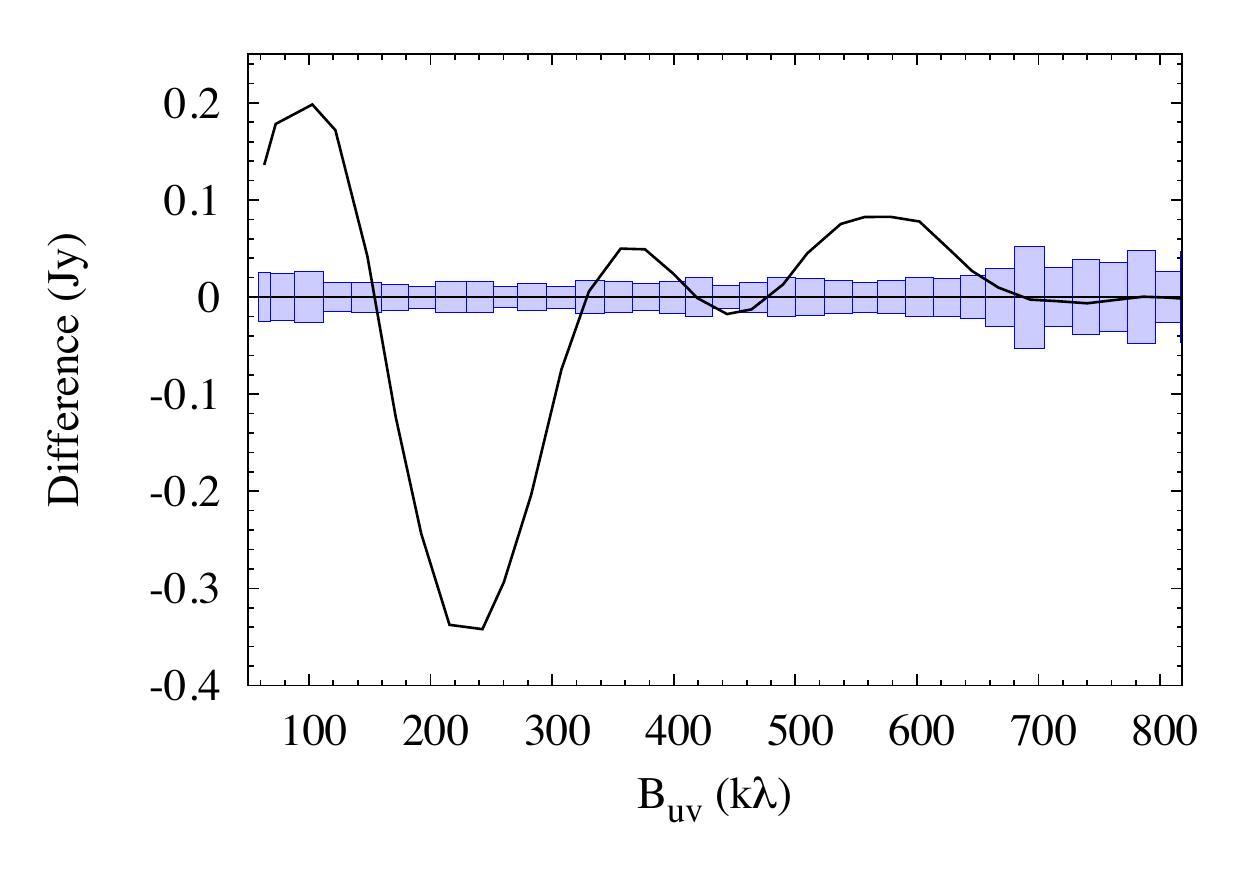}
\caption{\label{fig:alma2} Difference in the visibility profile
between the power-law and smooth viscous best-fit models (solid curve), compared to 
the 3$\sigma$ sensitivity of ALMA early science observations at 0.44~mm (shaded region). This figure demonstrates 
that ALMA observations can reliably distinguish between two plausible geometries for transitions disks 
to identify the mechanism likely responsible for inner cavities.
}
\end{figure}

Several experiments can be set up to investigate the origin of the inner cavities 
in transition disks by taking advantage of the superb imaging capabilities of the Atacama 
Large Millimeter Array (ALMA). Here we explore the possibility to use relatively low
angular resolution observations,  achievable during ALMA early science, to constrain 
the profile of the dust distribution at the edge of the dust 
continuum cavity. These observations can be used to disentangle the two possible scenarios
discussed above. As shown in Figure~\ref{fig:comparison}, and 
discussed in Section~\ref{sec:res_sigma}, CARMA observations fail to distinguish 
between a sharp and smooth edge of the continuum cavity because they do not
provide enough sensitivity on the spatial frequencies where the difference between 
the two models is the largest, namely between 100 and 700 k$\lambda$. 
On the other hand, Figure~\ref{fig:alma} shows that constraining the radial density profile at the 
edge of the continuum cavity is within the capabilities of ALMA in the early science 
phase. The left-most panels of Figure~\ref{fig:alma} show the model of the dust emission corresponding 
to the best-fit solutions for the smooth viscous and power-law surface density profile discussed in 
Section~\ref{sec:res_sigma}.  The maps in the right-most panels represent ALMA band 9 ($\lambda = 0.44 \mu$m) 
simulated observations obtained using the array specifications for Cycle 0 observations and 30 min 
on source integration. The structural differences between the two models are evident in 
these images:  the smoothly varying $\Sigma(r)$ shows trace amounts of emission in 
the inner disk that is detectable with ALMA, while the inner region in the truncated disk 
model is devoid of emission and resolved. More quantitatively, Figure~\ref{fig:alma2} 
shows the difference between the visibility profiles of the two different models as 
a function of the spatial frequency, where the colored region indicate the 3$\sigma$ measurement 
uncertainties from the simulated observations.  At the wavelength adopted for the simulation, 
the two models can be disentangled at high significance level on almost all spatial 
frequencies shorter than 600 k$\lambda$, which correspond to baselines shorter than 270 m.

\subsection{The LkCa~15 outer disk structure}
\label{sec:dis_radius}
The discrepancy in the outer radius of the dusty 
and gaseous disk remains an unsolved issue since, as discussed in 
Section~\ref{sec:res_sigma}, it cannot be reconciled by adopting an 
exponentially tapered surface density profile. A caveat to this analysis is that our
observations constrain the dust density only between 40-120 AU. 
The extrapolation of $\Sigma$ to larger radii relies therefore on the 
assumption that the prescription of Equation~\ref{eq:sigma_used} is valid across 
the all disk radial extent. 

To test this assumption we compare our results with the gas density inferred 
from the analysis of the $^{13}$CO (1-0)
emission by \cite{Pietu07}.  In LkCa~15, the 
$^{13}$CO (1-0) line is nearly optically thin and traces the density in the 
disk mid-plane. By comparing the observations to a power-law disk model, 
\cite{Pietu07} find that the $^{13}$CO surface density decreases as 
$r^{-3/2}$ from about $1.4\times10^{16}$ cm$^{-2}$ at 100 AU to 
$1.2\times10^{15}$ cm$^{-2}$ at 500 AU.  If the $^{13}$CO abundance
is constant throughout the disk, then the decrease of the 
gas density is less steep than what is predicted by an exponentially tapered 
profile.  

\begin{figure}[!t]
\centering
\includegraphics[angle=0, width=\columnwidth]{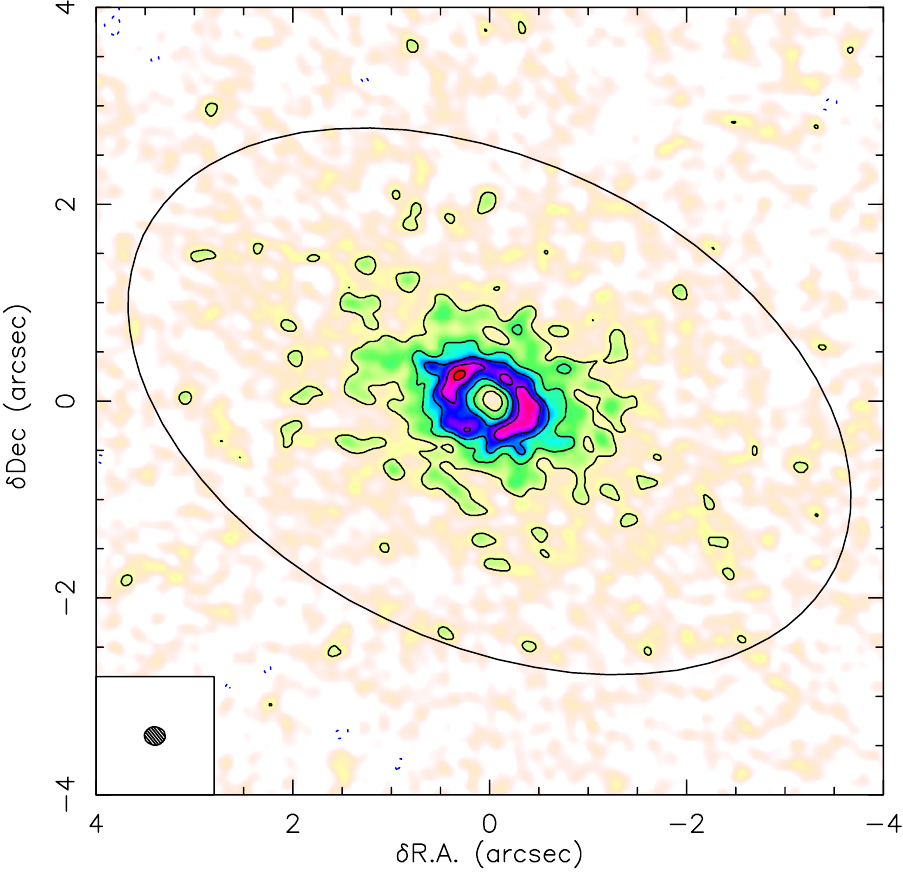}
\caption{\label{fig:pietu} Simulated observations of the 1.3 mm dust continuum emission 
corresponding to the surface density profile derived from the analysis of the optically 
thin $^{13}$CO line emission \citep{Pietu07}. The simulation have been obtained 
by assuming the same {\it uv} coverage and noise level as in the CARMA observations. 
The image as been made by adopting uniform weighting to achieve a resolution 
of 0.21\arcsec$\times$0.19\arcsec, as in panel (c) of Figure~\ref{fig:obs}. Contours
are plotted every 3$\sigma$, where the 1$\sigma$ noise level is 0.43 mJy/beam.
}
\end{figure}

To check whether the disk surface density profile derived from CO observations 
is consistent with the  dust continuum emission, we show in Figure~\ref{fig:pietu} 
simulated CARMA observations of the 1.3 mm dust emission corresponding to a dust surface 
 density $\Sigma_d(r) =  0.13 \times (r/100 \textrm{AU})^{-3/2}$ that extends 
 between 42.5 AU and 550 AU. The surface density normalization is derived 
 in order to reproduce the peak flux measured in our map. To facilitate the 
 comparison with the observations, the simulated emission is observed at the 
 same resolution and is affected by the same noise as in panel (c) of Figure~\ref{fig:obs}. 

We find that  only the dust emission within about 200 AU would have been clearly 
detected while the flux arising from larger radii would have been lost in the noise. 
The limited sensitivity of the continuum observations seems therefore to be 
the major responsible for the different radial extent of the dust and gas 
emission. Nevertheless, the fact that the simulated emission is about 
60\% more extended than what is observed suggests that either the 
dust-to-gas ratio or the dust opacity decreases at about 130 AU from the 
central star.

\begin{figure}[!t]
\centering
\includegraphics[angle=0, width=\columnwidth]{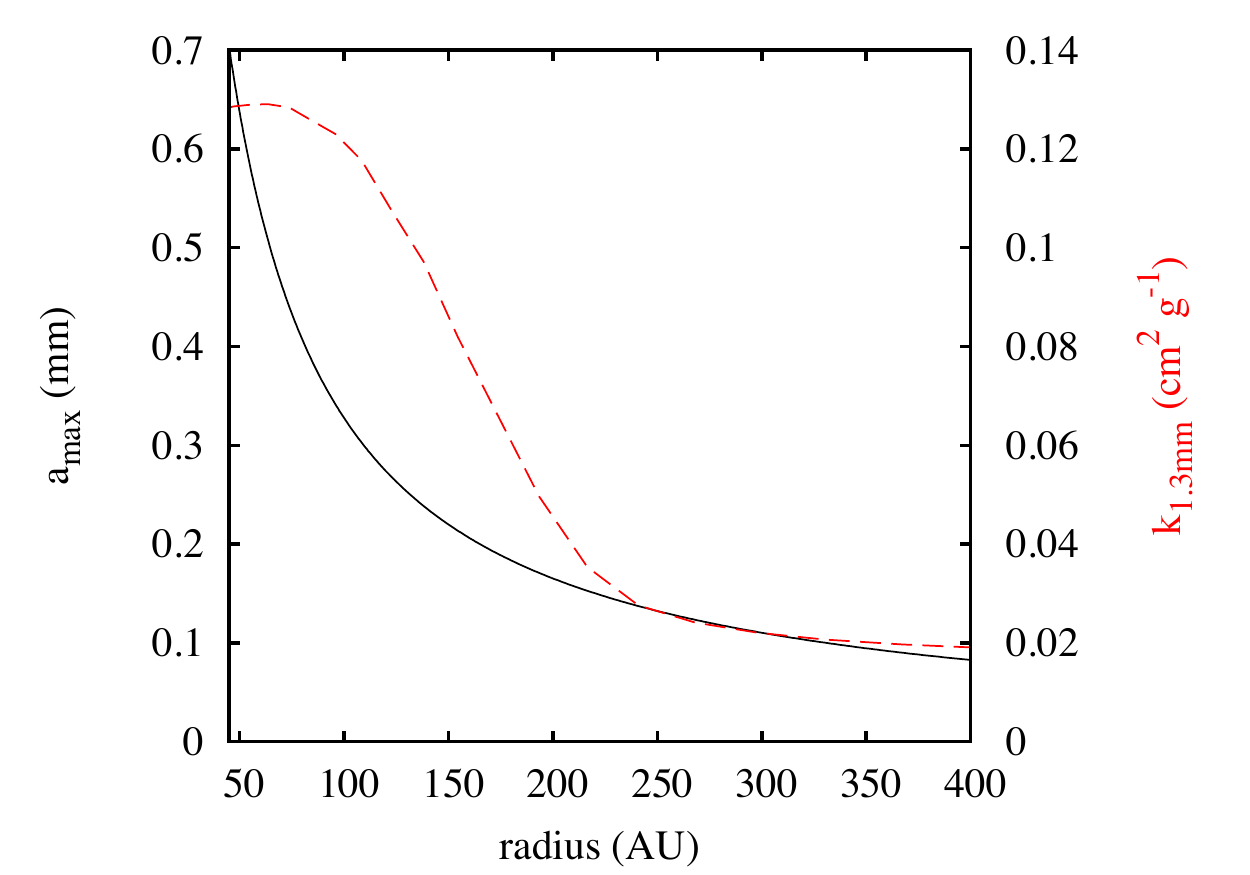}
\includegraphics[angle=0, width=0.7\columnwidth]{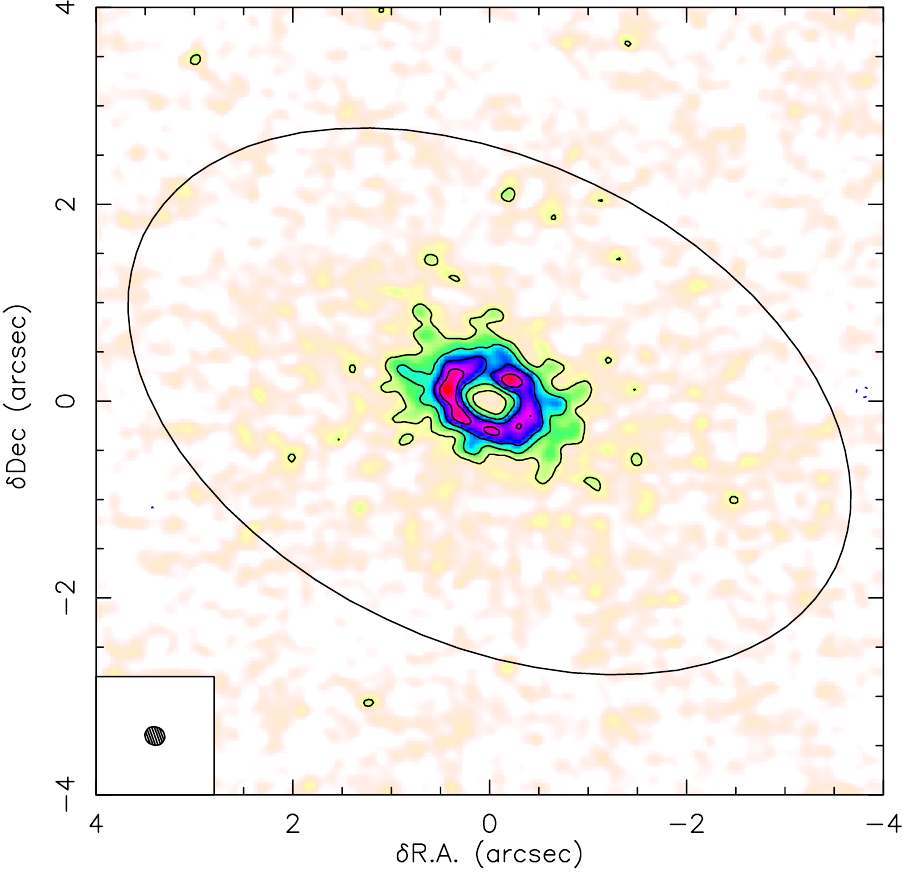}
\caption{\label{fig:amax} {\it Top panel:} Dependence of the maximum grain size $a_{max}$ (solid curve) and 
of the dust opacity $\kappa$ at 1.3 mm (dashed curve) on the orbital radius, calculated assuming 
the grain growth model discussed in Section~\ref{sec:disc}. {\it Bottom panel:} Simulated observations 
of the 1.3 mm dust continuum emission corresponding to the surface density profile derived from 
the analysis of the optically thin $^{13}$CO line emission \citep{Pietu07}, accounting for the 
radial variation of the dust opacity due to grain growth.  The decrease in the dust opacity 
between 100-200 AU results in a smaller effective radius for the dust 
continuum emission, compared to the case in which the dust opacity is constant throughout 
the disk (Figure~\ref{fig:pietu}).}
\end{figure}

In Section~\ref{sec:disc_inner} we suggested that grain growth is required 
to explain the structure of the innermost disk regions independently from 
the presence of planets. Here we argue that grain growth might also 
provide a natural explanation for the smaller extent of the dust continuum emission.
Grain growth models indeed predict 
that the outermost disk regions should be populated by  
sub-micron sized grains, which 
have a 1.3 mm dust opacity almost a factor 10 smaller than the mean dust opacity 
adopted in our disk modeling (Figure~\ref{fig:beta}). In the assumption that 
particles grow through mutual collisions and stick together by van der Walls forces, 
the maximum grain size $a_{max}$ of the grain size distribution can be expressed as \citep{Birnstiel10}
\begin{equation}
a_{max}(r) \simeq \frac{2\Sigma(r)}{\pi\alpha\rho_s}\frac{u_f^2}{c_s(r)^2} 
\end{equation}
where $\alpha$ is the viscosity parameter, $u_f$ is the dust fragmentation velocity, $\Sigma$ is 
the disk surface density,  $c_s$ is the sound speed , and $\rho_s$  the dust density. Figure~\ref{fig:amax}
shows the radial profile of $a_{max}$, and of the corresponding dust opacity at 1.3 mm, calculated by 
assuming  $\rho_s=2$ g cm${^2}$, $\alpha = 10^{-3}$, $u_f = 1$ m s$^{-1}$. 
The surface density is that derived from the CO observations as discuss above, and the sound speed 
$c_s$ is self-consistently calculated by assuming hydrostatic equilibrium  \citep[see, Appendix E in][]{Isella09}.
Although the exact values of $a_{max}$ depend on the assumptions on $\alpha$ and $u_f$, 
we find in general that a radial decrease of the maximum grain size corresponds to a drop in the 1.3 mm 
dust opacity. In particular, the decrease in $a_{max}$ from 0.4 mm to 0.1 mm between 
100 and 300 AU corresponds to a factor 5 decrease in the dust opacity.
Given the sensitivity of our observations, the drop in the opacity would in practice 
lead to a smaller radius of the continuum emission, making the simulation much alike 
the observations (bottom panel of Figure~\ref{fig:amax}).

This result suggests that the small disk radius measured
in the dust continuum emission might be the result of a combined decrease in dust opacity 
due to radial variations in the grain size distribution and the finite sensitivity of our observations. 
Investigating this hypothesis requires high-angular resolution multi-wavelength observations 
of the dust continuum emission.

\section{Summary}
\label{sec:summary}

We report new CARMA 1.3 mm dust continuum observations toward LkCa~15 which 
resolve the disk on spatial scales of 30 AU. We find that 90\% of the observed 1.3 mm dust 
emission arises from an azimuthally symmetric ring that contains about $5\times10^{-4}$ M$_{\sun}$ 
of dust. The remaining 10\% of the emission comes from a low-surface 
brightness tail that extends toward the northwest out to a radius of about 300 AU. 

To investigate the origin of the inner cavity observed in the dust emission we constrain the 
radial profile of the dust density at the edge of the cavity and across the dusty ring. 
The comparison with theoretical disk models shows that the surface density is rather flat 
between 40 and 120 AU, and that the drop of dust emission is consistent with either a sharp truncation of the dusty 
disk at 42.5 AU, or with a smooth inward decreases of the dust density between 3-85 AU. 
Within 40 AU, the observations constrain the amount of dust between $10^{-6}$ and 7 M$_\oplus$,  
where the minimum and maximum limits are set by the near infrared SED modeling and by the 
millimeter-wave observations of the dust emission respectively.

The recent discovery by \cite{Kraus11} of a  possible giant planet orbiting 
at 16 AU around LkCa~15 suggests that the dust distribution might be shaped by the 
dynamical clearing of this planet. However,  at least 3 other Jupiter size planets 
in addition to the one discovered by Kraus \& Ireland would be required to explain a 
sharp truncation in the dusty disk at 42 AU \citep{Zhu11,Dodson11}. In addition, 
the presence of planets should be coupled with low disk viscosity ($\alpha<5 \times 10^{-3}$) 
and with a significant depletion of micron-size grains, to explain both the presence of 
gas within the dust depleted cavity and the lack of near- and mid-infrared emission 
\citep{Zhu11}.  Alternatively, our 
results are consistent with a smooth inward decrease in the dust opacity caused by 
the coagulation of small grains into larger bodies, as predicted by grain growth models.
In this case, the dynamical clearing by the discovered planet would play a minor role in 
shaping the disk emission and no additional planets are necessary. 
We show that early science ALMA observations will be able to disentangle these 
two scenarios by constraining the slope of the 
dust density at the edge of the continuum cavity.

Finally, we find that 90\% of the dust emission is  confined within a radius of about 120 AU, while the gas 
emission is observed up to a radius of 900 AU. We find that this discrepancy cannot be 
reconciled using a exponentially tapered surface density profile but suggests a drop in 
the dust-to-gas ratio, or in the dust opacity, at radii larger than 120 AU. Using the grain growth 
model from \cite{Birnstiel10}, we show that a decline in the maximum grain size with radius 
will reduce the dust opacity by a factor of 5 between 100 and 300 AU, with a corresponding 
decrease in the dust continuum emission.
 
\vspace{0.5cm}

\noindent {\bf Note added in proof:\\}
Since the work described in this paper was completed and submitted for publication, 
850~$\mu$m observations of the dust continuum emission toward LkCa~15 that 
achieve a resolution of about 0.25\arcsec\ have been published by \cite{Andrews11b}.
These observations show that the continuum emission arises from a ring with the same 
properties of that discussed in this paper. In particular, they also find that the ring requires a 
rather flat surface density profile between about 40 and 120 AU.\\

\vspace{0.2cm}

\acknowledgments

We thank the OVRO/CARMA staff and the CARMA observers for their assistance 
in obtaining the data. Support for CARMA construction was derived from the Gordon and 
Betty Moore Foundation, the Kenneth T. and Eileen L. Norris Foundation, the James S. McDonnell 
Foundation, the Associates of the California Institute of Technology, the University of Chicago, 
the states of California, Illinois, and Maryland, and the National Science Foundation. Ongoing 
CARMA development and operations are supported by the National Science Foundation under 
a cooperative agreement, and by the CARMA partner universities.
We acknowledge support from the Owens Valley Radio Observatory, 
which is supported by the National Science Foundation through grant AST 05-40399.
This work was performed in part under contract with the Jet Propulsion Laboratory 
(JPL) funded by NASA through the Michelson Fellowship Program. JPL is 
managed for NASA by the California Institute of Technology. L.M.P. acknowledges support 
for graduate studies through a Fulbright-CONICYT scholarship.

\appendix
\section{Stellar proper motion}
The proper motion for LkCa~15 has been measured by \cite{Roeeser10} 
comparing {\it USNO-B1-0}  \citep{Monet03} and {\it Two Micron All Sky Survey}
data \citep{Skrutskie06}, and by \citet[][US Naval Observatory Twin Astrograph project]{Zacharias09}.
The first measurement leads to $\mu_\alpha \cos\delta = 9.1\pm2.1$ mas yr$^{-1}$ and $\mu_\delta = -13.8\pm2.1$ 
mas yr$^{-1}$. The second to $\mu_\alpha \cos\delta = 9.6\pm2.2 $ mas yr$^{-1}$ and  $\mu_\delta = -13.3\pm2.2$.
The J2000 coordinates for epoch 2000 for LkCa~15 corrected for the proper motion are RA=69.824139\arcdeg\ 
and Dec=+22.350945\arcdeg\ \citep{Roeeser10}, and  RA=69.8241427\arcdeg\  and Dec=+22.3509412\arcdeg 
\citep{Skrutskie06}. The difference between the two catalogs is about 13 mas, and is consistent with the quoted astrometric 
errors (i.e, 15~mas in both right ascension and declination in \citealt{Roeeser10}, and 30 mas in \citealt{Zacharias09} respectively). 
Our CARMA observation have been corrected by assuming 
a mean proper motion of $\mu_\alpha \cos\delta = 9.35$ mas yr$^{-1}$ and $\mu_\delta = -13.55$ mas yr$^{-1}$. 
The J2000 coordinates corrected for the proper motion at the epoch of the observations are listed in Table~\ref{tab:obs}.

\begin{table}[!h]
\caption{\label{tab:obs}              }
\begin{tabular}{llll}
\hline
\hline
Date                 &  $\Delta$t &  RA & Dec  \\
	                 & (yr)                   & (hh:mm:sec) & (deg:min:sec) \\
\hline
27 Oct 2007    &  7.816	        & 04:39:17.799 & 22:21:03.289 \\
10 Dec 2008  &  8.938              & 04:39:17.800 & 22:21:03.274  \\
26 Nov 2010  &	 10.898            & 04:39:17.801 &  22:21:03.247 \\
30 Jan 2011   & 11.075            & 04:39:17.801 &  22:21:03.245 \\
\hline 
\end{tabular}
\tablecomments{Column (1): date of the observations. Column (2): time baseline from year 2000. Column (3,4): stellar 
coordinates at the epoch of the observation assuming a proper motion of 
$\mu_\alpha \cos\delta = 9.35$ mas yr$^{-1}$ and $\mu_\delta = -13.55$ mas yr$^{-1}$.}
\end{table}


\bibliographystyle{apj}
\bibliography{ref}

\end{document}